\documentclass[conference]{IEEEtran}
\IEEEoverridecommandlockouts
\usepackage{amsmath,amsfonts}
\usepackage{algorithmic}
\usepackage{algorithm}
\usepackage{array}
\usepackage[font=normalsize,labelfont=sf,textfont=sf]{subcaption}
\usepackage{textcomp}
\usepackage{stfloats}
\usepackage{url}
\usepackage{verbatim}
\usepackage{graphicx}
\usepackage{enumitem}
\usepackage{xcolor}
\usepackage{bbm}
\usepackage{amsthm}
\usepackage{multirow}
\usepackage{svg}
\usepackage{gensymb}
\newfont{\bbb}{msbm10 scaled 700}

\DeclareSymbolFont{bbsymbol}{U}{bbold}{m}{n}

\newfont{\bb}{msbm10 scaled 1100}

\DeclareMathSymbol{\CC}{\mathbin}{bbsymbol}{'103}
\DeclareMathSymbol{\PP}{\mathbin}{bbsymbol}{'120}
\DeclareMathSymbol{\RR}{\mathbin}{bbsymbol}{'122}
\DeclareMathSymbol{\QQ}{\mathbin}{bbsymbol}{'121}
\DeclareMathSymbol{\ZZ}{\mathbin}{bbsymbol}{'132}
\DeclareMathSymbol{\FF}{\mathbin}{bbsymbol}{'106}
\DeclareMathSymbol{\GG}{\mathbin}{bbsymbol}{'107}
\DeclareMathSymbol{\EE}{\mathbin}{bbsymbol}{'105}
\DeclareMathSymbol{\NN}{\mathbin}{bbsymbol}{'116}
\DeclareMathSymbol{\KK}{\mathbin}{bbsymbol}{'113}
\DeclareMathSymbol{\HH}{\mathbin}{bbsymbol}{'110}
\DeclareMathSymbol{\SSS}{\mathbin}{bbsymbol}{'123}
\DeclareMathSymbol{\UU}{\mathbin}{bbsymbol}{'125}
\DeclareMathSymbol{\VV}{\mathbin}{bbsymbol}{'126}
\DeclareMathSymbol{\XX}{\mathbin}{bbsymbol}{'130}
\DeclareMathSymbol{\BB}{\mathbin}{bbsymbol}{'102}

\DeclareMathSymbol{\yy}{\mathbin}{bbsymbol}{'171}
\DeclareMathSymbol{\xx}{\mathbin}{bbsymbol}{'170}
\DeclareMathSymbol{\zz}{\mathbin}{bbsymbol}{'172}
\DeclareMathSymbol{\sss}{\mathbin}{bbsymbol}{'163}
\DeclareMathSymbol{\rr}{\mathbin}{bbsymbol}{'162}
\DeclareMathSymbol{\pp}{\mathbin}{bbsymbol}{'160}
\DeclareMathSymbol{\qq}{\mathbin}{bbsymbol}{'161}
\DeclareMathSymbol{\ww}{\mathbin}{bbsymbol}{'167}
\DeclareMathSymbol{\hh}{\mathbin}{bbsymbol}{'150}
\DeclareMathSymbol{\uu}{\mathbin}{bbsymbol}{'165}
\DeclareMathSymbol{\vvv}{\mathbin}{bbsymbol}{'166}

\DeclareMathSymbol{\indicator}{\mathbin}{bbsymbol}{'061}

\usepackage[mathscr]{euscript}

\usepackage{stmaryrd} % for \mkv 

\DeclareOldFontCommand{\rm}{\normalfont\rmfamily}{\mathrm}
\DeclareOldFontCommand{\sf}{\normalfont\sffamily}{\mathsf}
\DeclareOldFontCommand{\tt}{\normalfont\ttfamily}{\mathtt}
\DeclareOldFontCommand{\bf}{\normalfont\bfseries}{\mathbf}
\DeclareOldFontCommand{\it}{\normalfont\itshape}{\mathit}
\DeclareOldFontCommand{\sl}{\normalfont\slshape}{\@nomath\sl}
\DeclareOldFontCommand{\sc}{\normalfont\scshape}{\@nomath\sc}

\newcommand{\GREEN}{\color[rgb]{0,0.80,0.20}}

\usepackage[font={small}]{caption}

\allowdisplaybreaks

\usepackage{cite}
\usepackage[export]{adjustbox}

\usepackage{soul}

\usepackage[left=.635in,right=.635in,top=.7in,bottom=1.02in]{geometry}

\begin{document}
\bstctlcite{IEEEexample:BSTcontrol}

% \title{\huge MODRAD-SC -- the world's first O-RAN Radio Unit 
% combined with a Software-Defined Radio
% }
\title{\huge A Modular O-RAN Testbed Based on SRS Open Source O-CU/O-DU and Massive Beams Modular O-RU
}
\author{
	\IEEEauthorblockN{Fabian G\"ottsch\IEEEauthorrefmark{1}, Oriol Font-Bach\IEEEauthorrefmark{2}, 
    Andreas Benzin\IEEEauthorrefmark{1}, 
    Dennis Osterland\IEEEauthorrefmark{1}, 
    Andre Puschmann\IEEEauthorrefmark{2}, \\
    Felix-Christopher Lutz\IEEEauthorrefmark{1}, 
    Wilhelm Keusgen\IEEEauthorrefmark{3}, 
    Giuseppe Caire\IEEEauthorrefmark{3}
    }
\IEEEauthorblockA{\IEEEauthorrefmark{1}Massive Beams, Berlin, Germany}
\IEEEauthorblockA{\IEEEauthorrefmark{2}SRS, Barcelona, Spain}
\IEEEauthorblockA{\IEEEauthorrefmark{3}Technical University Berlin, Berlin, Germany}

\IEEEauthorblockA{E-mail: \{fabian.goettsch, andreas.benzin, dennis.osterland, felix.lutz\}@massivebeams.com \\
\{oriol.font, andre.puschmann\}@srs.io
\\
\{wilhelm.keusgen, caire\}@tu-berlin.de}
% \vspace{-.07cm}
}
  
\maketitle
\begin{abstract}
In this paper, we present a modular open radio access network (O-RAN) consisting of the 5G Core, a central (O-CU) and distributed  unit (O-DU) by Software Radio Systems (SRS) and an O-RAN radio unit (O-RU), MODRAD-SC, by Massive Beams (MB). OCUDU provides an open source 5G-compliant O-CU and O-DU solution developed by SRS, while MB's radio unit is a fully O-RAN compliant category A O-RU. According to O-RAN split 7.2a, OCUDU performs higher layer functions up to the high physical (PHY) layer, while the O-RU handles low PHY and RF functions. This results in an O-RAN-compliant 5G gNodeB. In an alternative configuration, OCUDU and MODRAD-SC operate in a software-defined radio fashion corresponding to split 8, facilitating non-real-time experiments among others. 
In both cases, the system provides full control over O-CU, O-DU, and O-RU. In addition, we will discuss the possibility to attach an analog beamformer to the O-RU, enabling hybrid digital-analog beamforming. The flexibility and modularity offered by OCUDU and MODRAD-SC enable the practical realization of a multitude of applications, ranging from 5G demonstrators to pre-6G experiments. The system addresses the requirements of academia and industry and is well-suited as an easy-to-use platform for experimental and practical deployments.
\end{abstract}


\section{Introduction}
In recent decades, wireless communication networks have increasingly become an indispensable part of everyday infrastructure. With every new generation of cellular or WLAN technology, parts of the standards have been changed to improve network performance. In parallel, huge progress in hardware and processing power have been achieved. The interaction of these advances has led to great improvements of data rates, latency, and coverage in wireless communication. 
These advancements have been obtained by theoretical and practical research, with some technologies and ideas developed in theory being deployed in practice many years or decades later, such as multiple-input-multiple-output (MIMO) wireless communication \cite{marzetta2016fundamentals}. In order to bring ideas from theory into practice, test environments and demonstrators are required before large-scale implementation can take place. For this, readily accessible test equipment is essential. In particular, it is beneficial if the test equipment 1) can cover different technologies and use cases, 2) is easy to configure and use, and 3) is affordable for companies, universities and research institutes. 

% {\RED [Maybe the next paragraphs are a bit too long. We can shorten this later.]}

For the architecture of next generation mobile communication systems such as 6G, Open Radio Access Network (O-RAN) systems have emerged as a promising approach. O-RAN is a disaggregated RAN with open interfaces, midhaul and fronthaul between the network entities, where low physical (PHY) layer functions are handled by O-RAN Radio Units (O-RUs) and high PHY operations by O-RAN Distributed Units (O-DUs) \cite{polese2023understanding}. This is an important difference to closed systems from previous mobile communication generations and enables mobile network operators (MNOs) to choose from a much wider range of network components, such as DU and RU providers. On the other hand, companies can focus on offering specific parts, software and hardware, of the network. The interoperability of equipment vendors leads to multi-vendor competition, driving innovation and optimization of, e.g., hardware cost and energy efficiency. 

An end-to-end O-RAN system consists of radio controllers, an O-RAN central unit (O-CU), O-DU and O-RU. It is connected to the core network on one side and to the user equipments (UEs) on the other side.
% \footnote{We refer the reader to \cite{polese2023understanding} for a description of the radio controllers and the core, and their tasks.} 
% The tasks of the core network include managing user connections, sessions, and data traffic, handling authentication, mobility, policy enforcement, and enabling advanced features like network slicing and edge computing. The near-real-time RIC performs near-real-time (in the order of milliseconds) control and optimization of O-RAN elements such as load balancing and interference management, while the non-real-time RIC handles the RAN policy management and performs non-time-sensitive optimizations (in the order of seconds). 
The O-CU is split into two logical components for the user (O-CU-UP) and control (O-CU-CP) plane, enabling different functionalities to be deployed in various locations across the network and on different hardware platforms. The O-DU is responsible for the radio link control (RLC), medium access control (MAC), and the higher part of the PHY, while the O-RU handles low PHY and RF functions. The specific distribution of PHY functions depends on the chosen split option, ranging from split 8 (only RF functions at the O-RU, FFT/IFFT (low-PHY) in the DU) to split 6 (O-RU handles the complete PHY).
The UE can be a commercial-off-the-shelf device or a measurement device for more advanced investigations in a testbed. A more detailed description of the network components (including the core and radio controllers) and their tasks can be found in \cite{polese2023understanding, agarwal2025open}

\subsection{Existing Open RAN and SDR Platforms}
OpenAirInterface{\texttrademark} (OAI) \cite{kaltenberger2020openairinterface} is an open-source software platform that implements 3GPP-compliant LTE (4G) and 5G protocol stacks for both the Radio Access Network (RAN) and core network elements. OAI can be used with different software-defined radios (SDRs), where USRP devices from ETTUS research are a common choice  \cite{kaltenberger2020openairinterface}. 
OAI is implementing the 5G stack in the OAIBOX{\texttrademark} solution offered by Allbesmart, which supports either research purposes using an SDR (i.e., O-RAN Split 8) or a real-time O-RAN system using a commercial, closed CAT-A O-RU (i.e., O-RAN Split 7.2a) \cite{oaibox_website}.
% OAI is implementing the 5G stack in the OAIBOX{\texttrademark} solution offered by Allbesmart, which also supports use of O-RUs and lab-based experimentation on FR2 (mmWave) and FR3 \cite{oaibox_website}. Another provider of O-RAN compatible and open-source O-DU and O-CU products is Airpuls \cite{airpuls_website_2026}, offering 5G solutions for a wide range of deployments such as research, testbeds, and private networks.

% {\RED [@Fabian, see if some reference from the removed paragraph needs to be rescued and added in the paragraph below for SDR/O-RU SoTA.] }

On the other side, several SDRs and O-RUs can be used in combination with the discussed O-CU/O-DU stacks. Software Radio Systems (SRS) has integrated a wide range of O-RUs with their 5G O-CU O-DU solution.
% The ``Eridan 5G Radio Unit 2TX 2RX'' is O-RAN compliant for split 7.2x and operates at carrier frequency $f_c \in [3.55, 3.7]~\text{GHz}$ with a bandwidth of 40~MHz and a transmit power of 0.16~W (22~dBm) per port \cite{eridan}. The ``5G Open Radio Unit White Box'' developed by Whizz Systems in collaboration with Intel, Analog Devices, Comcores, and Radisys supports 4T 4R TDD operation at $f_c \in [0.6, 6]~\text{GHz}$ with 100~MHz bandwidth \cite{WhizzSystems2024ORU}. Different frontends and SDR platforms for FR3 and mmWave frequency bands are developed by NYU spin-off Pi-Radio, used for hardware demonstrations and experimental research \cite{dhananjay2021pi, mezzavilla2024frequency}. 
Popular platforms include Ettus/National Instruments (SDRs), Liteon (sub-6 GHz and mmWave O-RUs), Benetel (sub-6 GHz O-RUs), Fujitsu (sub-6 GHz and mmWave O-RUs), and NEC (O-RUs). It is important to note that the CAT-A O-RUs are closed solutions and do not provide much flexibility, while SDR setups---often based on the USRP Hardware Driver (UHD)---face performance limitations (e.g., reduced bandwidth) when operating in real-time.

\begin{figure}[t!]
	\centering
	\includegraphics[width=\linewidth]{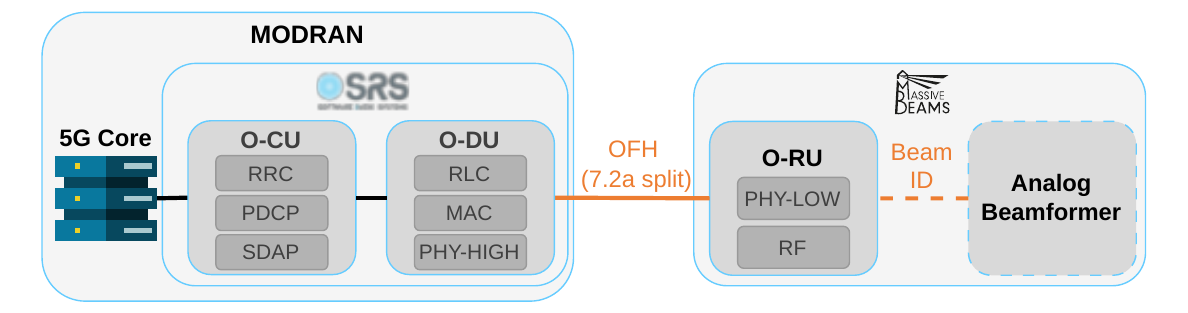}
	\vspace{-0pt}
	\caption{Overview of the SRS and MB system architecture with an external analog beamformer.}
	\label{fig_system}
    \vspace{-0pt}
\end{figure}
\subsection{Contributions of the OCUDU and MODRAD-SC Platforms}
The O-RAN system based on SRS' OCUDU, pre-installed on Massive Beams' (MB's) modular radio access network (MODRAN) stack, and MB's modular radio small cell (MODRAD-SC) addresses the requirements of both academia and industry for an accessible O-RAN system for testing and practical deployment.
OCUDU is an open-source carrier-grade 5G O-RAN O-CU and O-DU solution. The MODRAD-SC can be used as an O-RAN-compliant 5G base station, i.e., an O-RU platform for real-time experiments.
Combining the MODRAD-SC with OCUDU for unconstrained real-time experiments, as depicted in Fig. \ref{fig_system}, results in a high-performance O-RAN compliant \textit{modular radio access network}, with full control over O-CU, O-DU and O-RU.
The system depicted in Fig. \ref{fig_system} serves as a modular and open platform for both 5G O-RAN and 6G research and testbeds, where a wide variety of technologies can be investigated with custom frontend modules (FEMs).

The analog beamformer in Fig. \ref{fig_system} is optional in this setup, and will be discussed later in Section \ref{sec_system}. This system enables experiments and demonstrators for a wide range of technology and use case experiments, such as pre-6G trials, different split options and beamforming techniques in FR1 and FR3.
Additionally, the MODRAD-SC can be used as an SDR for non-real-time or performance-bounded real-time experiments involving non-standardized technologies and methods, like advanced waveform design and algorithm testing. OCUDU and MODRAD-SC can then be interfaced in a SDR fashion (split 8) to facilitate laboratory-based experimentation.
The flexibility of both elements, which conveniently supports a wide myriad of system configurations, thus makes its combination a versatile, high-performance and easy-to-use experimental platform.

In the next section, we will introduce OCUDU before presenting the MODRAD-SC platform in Section \ref{sec_modrad}. The combined OCUDU-MODRAD-SC system, potential use cases, and FR3 beamforming will be described in Sections \ref{sec_use_cases} and \ref{sec_system}, respectively.
%%%
%%%
%%%
%%%
\section{ Open CU DU (OCUDU) } \label{sec_srsran}
% {\RED [NOTE 1: Tentative OCUDU reference (to be updated).]}
%{\RED [NOTE 2: I didn't add reference to the O-RAN/3GPP specs we suppport (e.g., as done for MODRAD in O-RU mode), as they would be many. But I can do.]}
%{\BLUE [@Oriol: I think it makes more sense not to mention all the specs of the MODRAD in O-RU mode. It was a bit heavy and I reduced it. So, I think your part is already quite good in the current version.]}

OCUDU \cite{ocudu_git} provides a commercial-grade 5G O-CU/O-DU from SRS, taking the basis of srsRAN \cite{srsRAN} and growing it into a thriving open-source ecosystem with advanced capabilities in artificial intelligence (AI) and 5G/6G. OCUDU offers a complete RAN solution that complies with both 3GPP and O-RAN Alliance specifications. It includes the complete L1/2/3 stack, developed in-house and with minimal external dependencies. The software is portable across processor architectures, scalable from low-power embedded systems to cloudRAN, and is designed to support deployment on a diverse set of hardware platform configurations, including CPU/GPU combinations for accelerated compute support. It also features AI-native RAN capabilities developed by DeepSig, making it a very powerful platform for experimental-driven mobile wireless research and development. Moreover, OCUDU aims at aligning public and private goals across academia, government and industry to create the “Linux of RAN”, a foundational software solution for 5G, 6G and beyond.

\begin{figure}[t!]
	\centering
	\includegraphics[width=\linewidth]{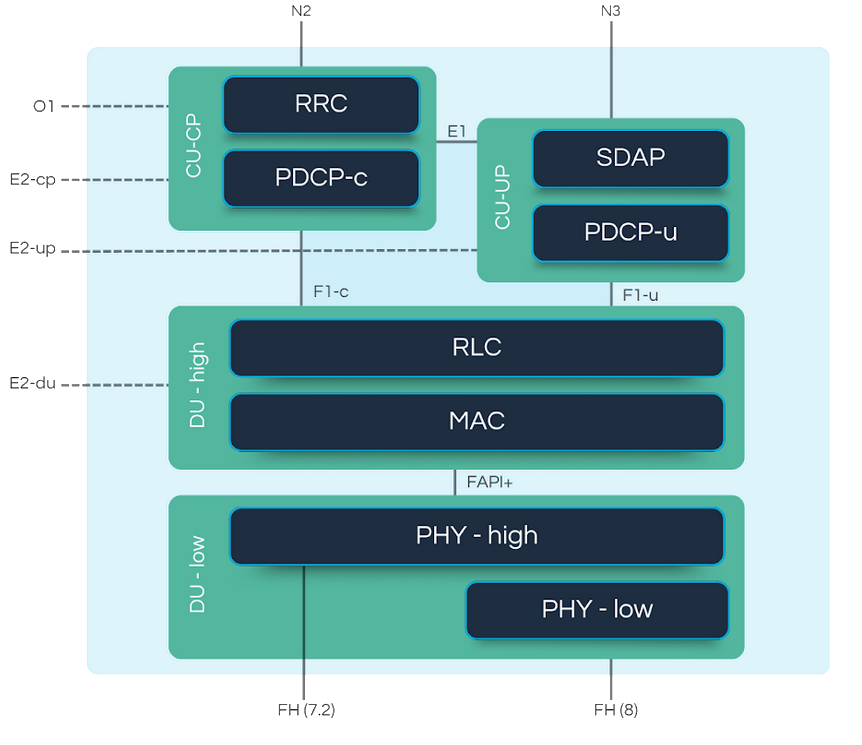}
	\vspace{-0pt}
	\caption{Overview of the O-CU/O-DU solution provided by OCUDU.}
	\label{fig_ocudu_diagram}
    \vspace{-0pt}
\end{figure}

The O-CU/O-DU solution from SRS includes the required 3GPP and O-RAN interfaces to enable disaggregated deployments and interact with third-party O-RAN equipment and AI-driven management entities, as shown in Figure \ref{fig_ocudu_diagram}. Split deployment of the O-CU and O-DU is possible through the F1 interface, whereas the E1 interface enables the O-CU-UP and O-CU-CP separation. More relevantly to the scope of this paper, the O-DU supports both split 8 to interface MODRAD-SC as a software defined radio, and split 7.2a when the latter operates as an O-RU (including M-plane support over the OFH interface). Furthermore, both the E2 and O1 interfaces enable AI-driven RAN monitoring and control by interfacing to the near-real-time RIC and to the non-real-time RIC (as part of the service and management orchestration (SMO) entity), respectively. Table \ref{table_ocudu_parameters} summarizes the main parameters currently supported by OCUDU's 5G RAN solution. Other significant features include support for GEO NTN, RAN slicing, multi-cell, positioning, and various types of handover (intra- and inter-CU, intra-DU, conditional, Xn-based).

\begin{table}[t!]
    \caption{Main parameters of OCUDU.}
    \begin{center}
    \begin{tabular}{ | p{.367\linewidth} | c | c |}
		\hline
		{\textbf{Parameter}} & {\textbf{Value}} \\ \hline
        Channel bandwidth [MHz] & 100\\ \hline
		Frequency bands  & FR1 (15/30 kHz), FR2 (120 kHz) \\ \hline
		Antenna ports \{Tx, Rx\} & { \{4, 4\} } \\ \hline
        Standard alignment & 3GPP (rel. 17), O-RAN \\ \hline
	\end{tabular}
	% \vspace{-1.0em}     
    \label{table_ocudu_parameters}
    \end{center}
\end{table}

It is also relevant to highlight that the features of OCUDU will grow steadily over the next two years. In this regard, current ongoing development includes support for \textit{O-RAN Alliance compliant CAT-B O-RUs} (split 7.2b), 8T8R, release 17 NR NTN (LEO/MEO), beam management and angle-based positioning for FR2, latency-bounding features towards ultra-reliable low-Latency communications (uRLLC) support (e.g., UL pre-scheduling and configured grant), as well as initial FR3 support in combination with the MODRAD-SC. Moreover, the planned roadmap includes very relevant extensions to pre-6G research, including support for multi-user MIMO, reprocity-based and hybrid beamforming, massive MIMO (up to 64T64R, combined with MB's massive MIMO MODRAD), release 18+ NR NTN and the inclusion of deeply-embedded stack hooks (in the form of APIs) towards physical-layer experimentation and real-time control-loop (dApps) support.

\section{MODRAD-SC Platform} \label{sec_modrad}
The MODRAD-SC is an in-house development by Massive Beams (MB) for use as O-RU and SDR. 
As a 5G-NR Open RAN compliant O-RU, MODRAD-SC supports all features of the Open RAN fronthaul interface and can be combined with a O-DU/O-CU solution.
In SDR mode, MODRAD-SC can operate in TDD and FDD mode in different bands covering FR1, FR2, and FR3 frequency ranges with the corresponding IF/BB interfaces. MODRAD-SC enables valuable insights into heterogeneous access scenarios, cross-band prototyping, and the seamless integration of SDR technology with Open RAN architectures. 

In addition, due to its combination of high computational power, real-time processing capabilities, and high-speed fiber connectivity, MODRAD-SC can be employed as a universal controller for external add-ons like, e.g., phased array systems, frequency extenders, or other front-end extensions. This makes MODRAD-SC well suited to drive beamformer operation while meeting stringent timing requirements of under 300 microseconds. Through these features, MODRAD-SC represents both a powerful 5G base station for Open RAN environments and a generic, future-oriented platform for phased array control and beamforming (BF) research. MB's roadmap includes includes a massive MIMO MODRAD supporting O-RAN split 7.2b (i.e., CAT-B O-RU), FR3 FEMs, and Giga-MIMO with up to 256 antenna elements.
Note that this will be the first open CAT-B O-RU and SDR platform supporting massive MIMO.
% The system performance can be validated and benchmarked using measurement equipment. 

\subsection{Open RAN O-RU Mode}
In the standard configuration, MODRAD-SC is a fully \textit{O-RAN Alliance compliant CAT-A O-RU} based on split 7.2a. It can be seamlessly integrated with major O-DU/O-CU software stacks from, e.g., SRS \cite{ocudu_git} to realize a fully 3GPP and O-RAN compliant 5G-NR gNodeB. Together with a third-party 5G Core, a complete 5G-NR radio access network can be created and commercial off-the-shelf 5G user equipments can be attached.
As an O-RAN small cell Cat-A O-RU, the MODRAD-SC operates at a carrier frequency between $3.3$ and $3.8$~GHz with a bandwidth of $200$~MHz, where 4 transmission and 4 receive antenna ports are used. Note that $200$~MHz of instantaneous bandwidth are achieved with two carriers and $100$~MHz bandwidth per component carrier.

The MODRAD-SC radio unit features an O-RAN compliant fronthaul interface based on split 7.2a, as specified in O-RAN.WG4.TS.CUS.0-R003-v11.00, while the fronthaul link is compliant with IEEE 802.3 Ethernet technology.
% with 25G PHY and MAC layers
% supporting MTU sizes of 1500 bytes and jumbo frames up to 9000 bytes. VLAN tagging and QoS are implemented according to IEEE 802.1Q standards, enabling robust traffic prioritization and network segregation.
% Furthermore, the MODRAD-SC supports both 16-bit uncompressed IQ data and 9-bit block floating point (BFP) compressed IQ data. 
Hence, the MODRAD-SC supports all mandatory features of the O-RAN defined fronthaul interface for the Control, User and Synchronization plane (CUS planes) to assure interoperability, scalability, and future-proof operation. In addition, MB's in-house developed modular management plane (M-Plane) is fully compatible with SRS' M-plane.
% It accommodates up to 24 concurrent streams, including 8 downlink (DL) eAxC-IDs, 8 uplink (UL) eAxC-IDs, and 8 UL PRACH eAxC-IDs.
% The low-PHY PDSCH and PUSCH processing is based on 4096-point IFFT/FFT, while the low-PHY PRACH supports long and short preamble formats.
% The O-RU delay category is ``P'', with a nominal minimum one-way delay value in the uplink direction between 51 and 70 microseconds. 
%
% The control plane (C-Plane) supports section types 0, 1, and 3, facilitating a comprehensive layer of management commands and signaling. The user plane (U-Plane) implements application layer fragmentation to efficiently manage large payloads and maintain packet integrity during transport.
%
% Synchronization and timing in the synchronization plane (S-Plane) are assured via IEEE 1588 Precision Time Protocol version 2 (PTPv2) and Synchronous Ethernet (SyncE). The S-Plane supports the ITU-T G.8275.1 profile and complies with O-RAN S-Plane Link Layer Synchronization options LLS-C1, -C2, and -C3 topologies, essential for precise and stable timing across distributed network elements.
%
% Management plane (M-Plane) functionalities include configuration and control via a user-friendly web interface or secure SSH access over the fronthaul link. 
%
% The MODRAD-SC follows the O-RAN Alliance O-RAN.WG4.CUS.0-v07.00 specification, supporting all mandatory O-RAN features to assure interoperability, scalability, and future-proof operation.

\subsection{Ethernet SDR mode}
When operated in Ethernet SDR mode, the MODRAD-SC will become a fully tunable and customizable SDR, which can stream arbitrary time-domain IQ samples from and to the on-system RAM of the MODRAD-SC compute module/baseband processor. Through this feature, arbitrary waveforms (e.g., for 6G candidate technology experiments) can be radiated and received through the 4 antenna ports. 
The IQ sample streaming can be realized via, e.g., MATLAB and Python, for which drivers and example scripts are available.
% will be shipped together with MODRAD-SC. 
In Ethernet SDR mode, the MODRAD-SC operates with up to 4 transmission and 6 receive antenna ports. Carrier frequencies in the range between $0.08$ and $18$~GHz with bandwidths of $200$~MHz (Tx) and $400$~MHz (Rx) are supported, where 8~GByte of Sample RAM for time-domain IQ sample playback are available. 

\begin{table}[t!]
    \caption{Parameters of MODRAD-SC in O-RU and SDR mode.}
    \begin{center}
    \begin{tabular}{ | p{.367\linewidth} | c | c |}
		\hline
		{\textbf{Parameter}} & {\textbf{O-RU mode}} & {\textbf{SDR mode}} \\ \hline
        Channel bandwidth [MHz] & 200 (Tx, Rx) & 200 (Tx), 400 (Rx) \\ \hline
		Carrier frequency $f_c$ [GHz] & \multicolumn{2}{c|}{ 0.08 - 18 } \\ \hline
        Antenna ports & \multicolumn{2}{c|}{  4 Tx, 4 Rx, 2 ORx } \\ \hline
        \multirow{2}{*}{Mean output power per port} 
            & \multicolumn{2}{c|}{27 dBm at $f_c \in [3.3, 3.8]$~GHz}  \\ \cline{2-3} 
            & \multicolumn{2}{c|}{5 dBm otherwise}    \\ \hline
	\end{tabular}
	\vspace{-1.0em}     
    \label{table_modrad_parameters}
    \end{center}
\end{table}
\begin{figure*}[t!]
	\centering
	\includegraphics[width=\linewidth]{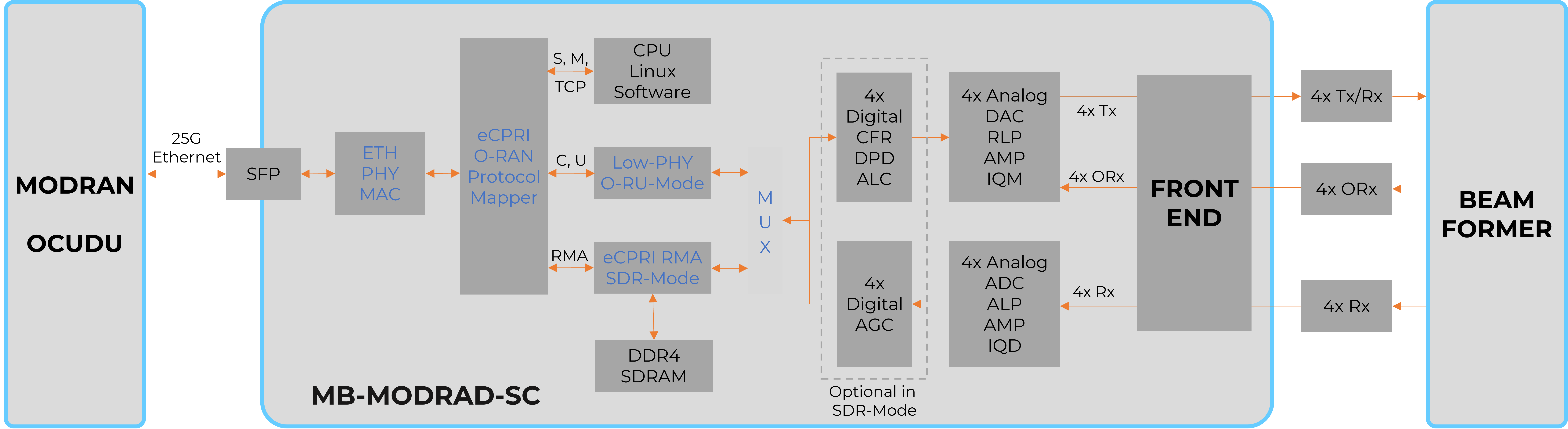}
	\vspace{-4pt}
	\caption{Detailed configuration of the MODRAD.}
	\label{fig_MODRAD_schematic}
    \vspace{-4pt}
\end{figure*}

\subsection{Components and Processing Blocks} \label{sec_blocks}
\begin{figure}[t!]
	\centering
	\begin{subfigure}{0.99\linewidth}
		\includegraphics[trim={11.8cm 12.2cm 4.3cm 22.3cm}, clip, width=\linewidth]{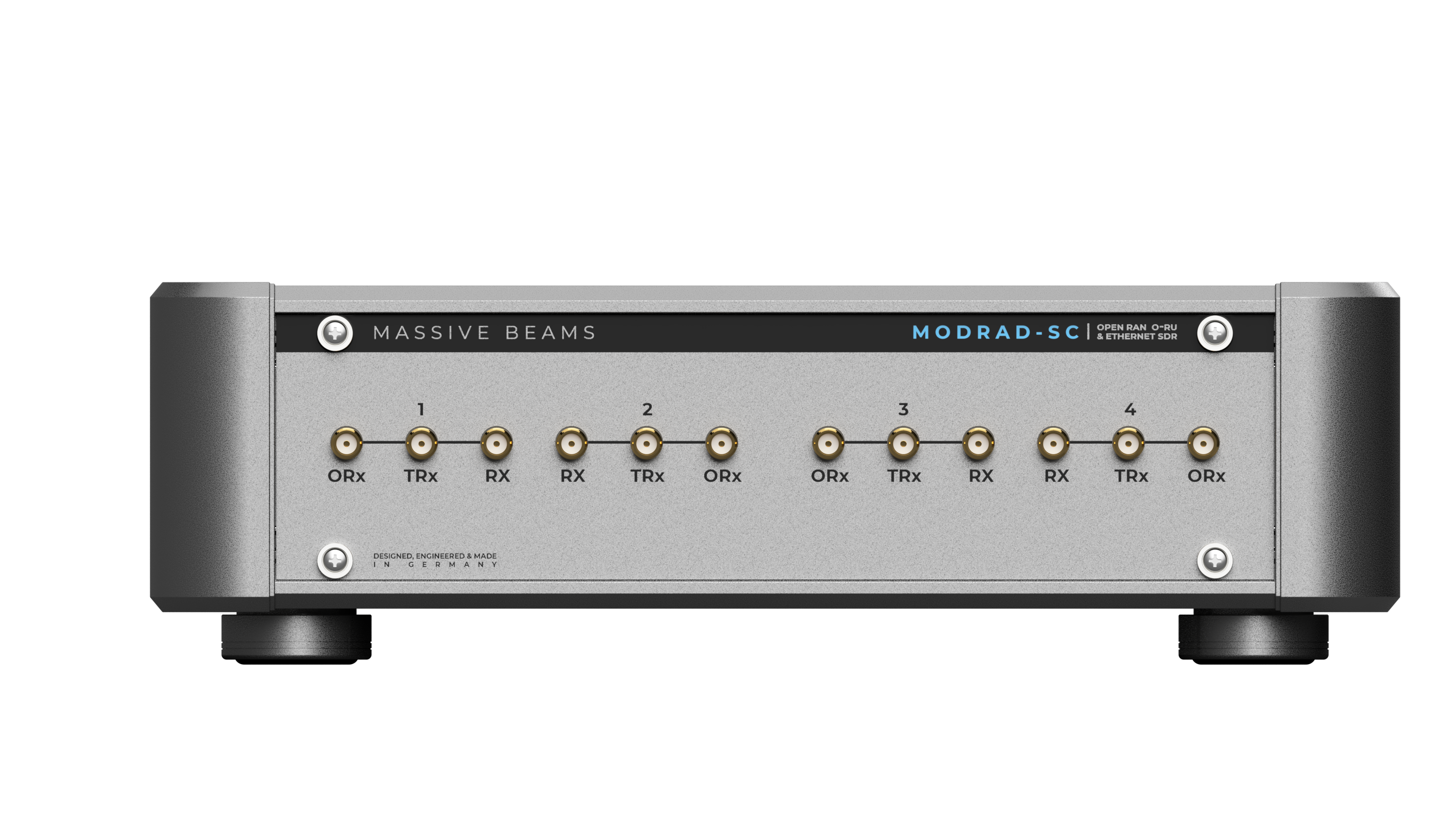}
	\end{subfigure}
	\begin{subfigure}{0.99\linewidth}
        \vspace{1pt}
		\includegraphics[trim={15.5cm 15.3cm 10.4cm 22.3cm}, clip, width=\linewidth]{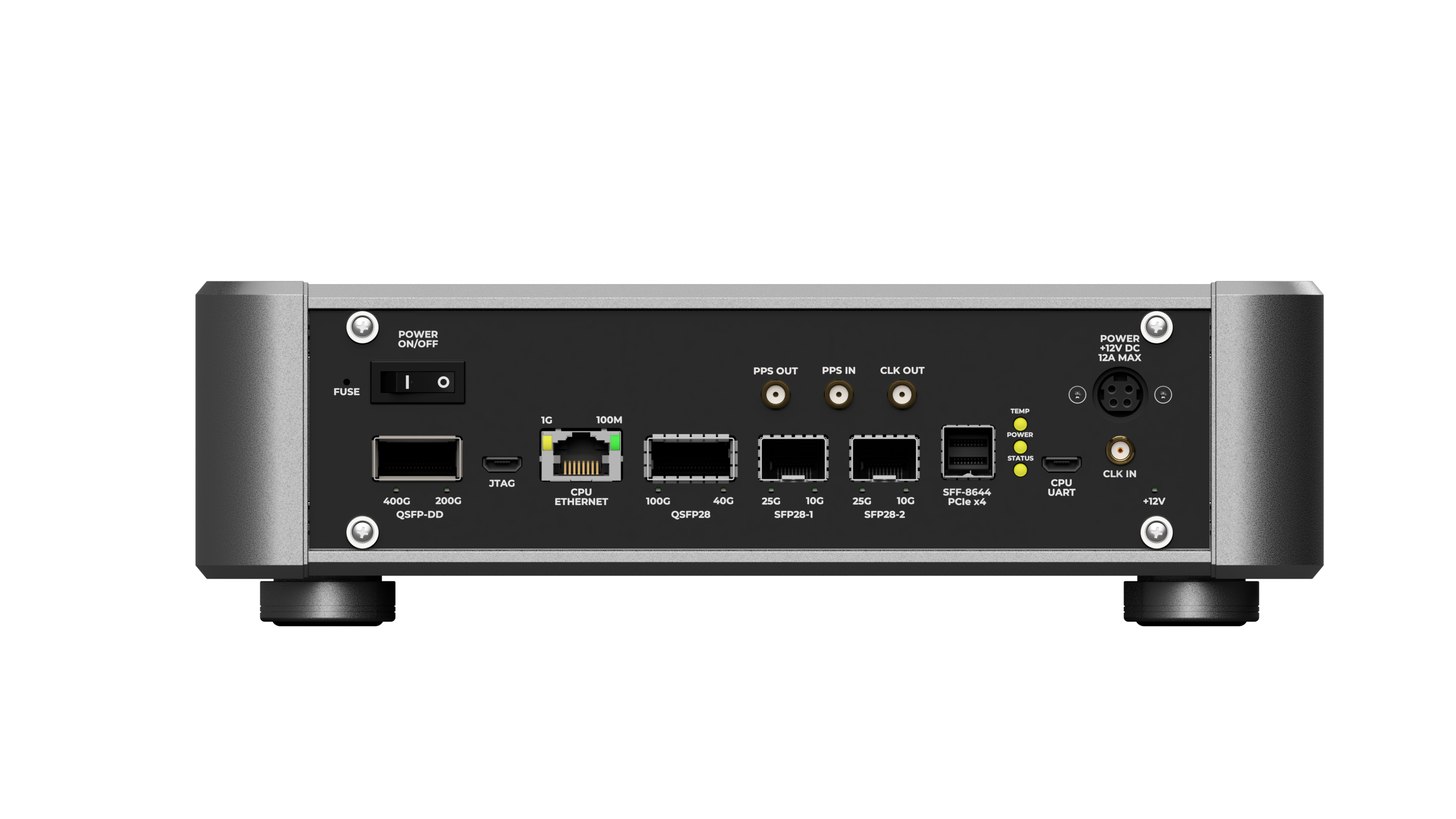}
	\end{subfigure}
    \vspace{-4pt}
	\caption{Front (top) and back (bottom) of the MODRAD.}
	\label{fig_MODRAD_front_back}
    \vspace{-4pt}
\end{figure}

This section describes the processing blocks and physical interfaces of the MODRAD-SC depicted in Figs. \ref{fig_MODRAD_schematic} and \ref{fig_MODRAD_front_back}, respectively.
\subsubsection{Processing Blocks} 
The processing blocks of the MODRAD-SC are shown in Fig. \ref{fig_MODRAD_schematic} and explained in the following:
\begin{itemize}[leftmargin=0pt]
    \item[] \textbf{``SFP''}: Small Form-factor Pluggable (SFP) transceiver, which is used to connect the MODRAD-SC to the O-DU/O-CU in O-RU mode and for IQ sample playback and recording in SDR mode. {\GREEN }
    \item[] \textbf{``ETH PHY MAC''}: The Ethernet (ETH) MAC (Medium Access Control) and PHY are part of the ETH switch that is connected to the O-DU/O-CU via the SFP; used for data transmission, frame handling, and interfacing with network protocols.
    \item[] \textbf{``eCPRI O-RAN Protocol Mapper''}: Converts Ethernet packets to Enhanced Common Public Radio Interface (eCPRI) and O-RAN compliant formats; parses and assembles fronthaul data according to O-RAN standards.
    \item[] \textbf{``CPU Linux Software''}: Runs the O-RU software stack and implements the O-RU configurations.
    \item[] \textbf{``Low-PHY O-RU-Mode''}: Implements lower PHY layer functions as defined for O-RUs, e.g., BF and IFFT/FFT.
    \item[] \textbf{``eCPRI RMA SDR-Mode''}: The Radio Modem Adapter (RMA) implements the SDR mode, providing direct adaptation of raw eCPRI payloads (I/Q samples, control signals) for flexible, software-driven signal processing.
    \item[] \textbf{``DDR4 SDRAM''}: Memory buffer for the eCPRI RMA in SDR mode, used to store I/Q samples and control data.
    \item[] \textbf{``MUX''}: Multiplexer to select and transmit multiple input signals through a single output line.
    \item[] \textbf{``4x Digital CFR DPD ALC''}: Digital blocks for crest factor reduction (CFR), digital pre-distortion (DPD), and automatic level control (ALC), improving PA efficiency and signal fidelity.
    \item[] \textbf{``4x Digital AGC''}: Automatic gain control (AGC) for optimization of varying input signal levels.
    \item[] \textbf{``4x Analog DAC RLP AMP IQM''}: Analog blocks for digital-to-analog conversion (DAC), reconstruction low-pass (RLP) filtering, amplification (AMP), and I/Q modulation (IQM), converting processed digital signals for transmission.
    \item[] \textbf{``4x Analog ADC ALP AMP IQD''}: Analog-to-digital conversion (ADC), aliasing low-pass (ALP) filtering, amplification (AMP), and I/Q demodulation (IQD) for incoming analog signals, enabling digital post-processing.
    \item[] \textbf{``FRONTEND''}: %{\BLUE [@Oriol: Can we link the following frontend module description better to the OCUDU and  potential use cases?]} 
    The last two blocks are connected to a FEM, which in turn is connected to the antennas. Note that MODRAD-SC can be used with a wide range of different custom FEMs, including the FR1, FR2, and FR3 bands, SDR-only mode, and based on given specifications for different technologies and use cases such as NTN. This flexibility is fully leveraged when integrated with OCUDU, enabling both real-world and lab-based experimentation, as already explained in Section \ref{sec_srsran}. In Sections \ref{sec_system} and \ref{sec_use_cases}, we will discuss some example use cases for a system combining OCUDU and MODRAD-SC.
\end{itemize}

% As an example we consider Fig. \ref{fig_MODRAD_schematic}, where the MODRAD-SC is connected to MB-FEMSC-n78-4T4R, a frontend for the n78 band (3.3 - 3.8~GHz). It enables TDD operation in the n78 band with two component carriers, each with 100~MHz of bandwidth and 4~Tx, 4~Rx for $4 \times 4$ MIMO operation where each of the $4$ antenna ports delivers a maximum of $+27$~dBm ($0.5$~W) mean output power. It consists of the following processing blocks and components:
% \begin{itemize}[leftmargin=*]
%     \item ``4x PA'': Four individual power amplifiers for each antenna port.
%     \item ``4x FBC'': Feedback coupler for monitoring PA output.
%     \item ``TDD DPLX'': Time division duplex (TDD) duplexer (DPLX) for separating signals in uplink and downlink directions.
%     \item ``MUX'': Multiplexer for combining processed paths before Rx/Tx antenna feed.
%     \item ``4x Tx/Rx'': Four direct conversion transmitters (Tx) or receivers (Rx), supporting $4 \times 4$ MIMO operation and spatial multiplexing.
%     \item ``4x ORx'': Two direct conversion observation receivers (ORx) with two switchable inputs each for feedback and calibration, especially useful for digital predistortion and diagnostics (only used in SDR mode).
%     \item ``4x Rx'': Four receive paths via Low-Noise Amplifiers (LNAs) to maximize incoming signal sensitivity before digitization (only used in SDR mode).
%     \item ``4x LNA'': Low noise amplifiers for each receive channel, minimizing noise and maximizing dynamic range of received signals.
% \end{itemize}
%%%%
%%%%
%%%%
%%%%
\subsubsection{Physical Interfaces}
The physical interfaces to the transmitter, receiver and observation antenna ports are 12 SMA (female) ports located on the front of the MODRAD-SC, as shown in Fig. \ref{fig_MODRAD_front_back} (top). 
The physical interfaces on the back of the MODRAD-SC are shown in Fig. \ref{fig_MODRAD_front_back} (bottom).  
They include an SMA (female) reference clock input and an SFP28 25G Ethernet interface used for 1) O-RAN C, U, S \& M plane fronthaul in O-RU-mode and 2) IQ sample playback and recording in SDR mode. Two HDMI High-Speed general-purpose input/output (GPIO) connectors, QSFP-DD and QSFP28, are available for accessory or external component control. For debugging purposes, the MODRAD-SC offers an RJ45 1G Ethernet, two Micro-USB Universal Asynchronous Receiver Transmitter (UART) and Joint Test Action Group (JTAG) connectors. 

% and explained in the following.
% \begin{itemize}[leftmargin=*]
%     % \item IEEE 802.3 Ethernet-compliant 25G PHY+MAC (Physical Layer and Media Access Control) with support for MTU=1500/9000 (Jumbo frames), VLAN, and QoS (IEEE 802.1Q) is used in small cell and software-defined radio (SDR) systems to provide high-speed, standards-based network connectivity between radio hardware and network infrastructure or centralized processing units.
%     \item 1x SFP28 25G Ethernet interface used for 1) O-RAN C, U, S \& M plane fronthaul in O-RU-mode and 2) IQ sample playback and recording in SDR mode
%     \item 2x HDMI High-Speed general-purpose input/output (GPIO) connectors, QSFP-DD and QSFP28, for accessory or external component control.
%     \item +12V DC via 6-pin PCI-E power connector
%     \item 1x RJ45 1G Ethernet for debugging purposes
%     \item 2x Micro-USB Universal Asynchronous Receiver Transmitter (UART) and Joint Test Action Group (JTAG) for debugging purposes
%     \item 1x SMA (female) reference clock input
% \end{itemize}
Notice that both the O-RU and the SDR modes rely on the same single 25 Gbit/s Ethernet link using one SFP28 pluggable transceiver. To change between O-RU and SDR mode no physical hardware changes or re-wiring needs to be done. All payload and control signaling is done through the single 25G Ethernet link, which is fully IEEE 802.3 Ethernet compliant and thus can be connected to any Ethernet switch or network interface card which supports 25G speeds.

% \begin{figure}[t!]
% 	\centering
% 	\includegraphics[width=\linewidth]{figures/RIS_main_picture.jpg}
% 	\vspace{-0pt}
% 	\caption{Feeder-RIS module {\color{red}[TBD: with the feeder array being connected to the MODRAD-SC ?]} in an anechoic chamber.}
% 	\label{fig_RIS}
%     \vspace{-0pt}
% \end{figure}

%%%
%%%
%%%
%%%
\section{Modular Open Radio Access Network} \label{sec_system}
In this section, we describe the platform resulting from the combined OCUDU and MODRAD-SC solutions. We will line out potential use cases and technologies that can be served and investigated with our system. In addition, we will describe in detail how this system can realize BF in the FR3 band in a fully O-RAN compliant way, using a generic beamformer (and in particular an NFED-RIS) attached to the MODRAD-SC.

\subsection{System Architecture}
The 5G Core is connected to the OCUDU, which in turn is connected to the MODRAD-SC (i.e., the O-RU), as shown in Fig. \ref{fig_system}. 
The O-CU's tasks include Radio Resource Control (RRC), Packet Data Convergence Protocol (PDCP), and Service Data Adaptation Protocol (SDAP). 
The O-DU hosts the radio link control (RLC), media access control (MAC), and high physical (High-PHY) layers of the 5G NR protocol stack, while the Low-PHY and radio frequency (RF) functions are performed at the O-RU. 
The MODRAD-SC is equipped with 4 RF chains that can be used for different types of BF strategies. While 4 transmit antennas may be sufficient for sub-6 GHz systems, for higher carrier frequencies a larger number of antennas is required to obtain a satisfying coverage and BF gain. In that case, the 4 RF chains at the MODRAD-SC can be used to steer or illuminate a BF array equipped with a larger number of (e.g., passive or metamaterial) antennas. 
The near-field fed RIS (NFED-RIS) developed by MB (see \cite{osterland2025evaluation, osterland2026evaluation, lutz2026design}) is a potential BF array to be used in combination with the described system. The hybrid digital-analog BF approach using this architecture in an O-RAN system will be explained in the following.

\subsection{Use Cases} \label{sec_use_cases}

%{\RED [To be done:] Add intersection points between MODRAD and OCUDU - e.g., NTN. Highlight that the combination of both produces a very powerful R\&D platform (as we will prove in 6G-LEADER and MULTIPLY-6G ;).) So far I've highlighted a couple. Based on this, perhaps we can expand the proposed use cases: FR1/FR3 coexistence, multi-user and massive mimo, etc.}{\BLUE [@Oriol: I tried to emphasize that we provide a powerful R\&D platform and to improve the description of some use cases. Please feel free to make changes and add text. Especially the use cases ``Multi-user and massive MIMO'', ``Sensing and localization'', ``NTN signaling'', ``FR1/FR3 coexistence'' may still be a bit too general. I also added ``AI-driven network management'' since we mention it in the conclusions. Feel free to add something, but we can also remove that point. From our side, it is not something we are currently working on.]}

Due to its modular nature, OCUDU and MODRAD-SC can support a wide range of experiments for different technologies and use cases.  
In addition, both platforms offer complete control over O-CU, O-DU, and O-RU, enabling a customized system for specific experimental and operational requirements.
This positions OCUDU and MODRAD-SC as a powerful research and development platform for both academia and industry. In the following, we will provide potential use cases in addition to the BF approaches discussed previously, focusing on promising research topics for upcoming 6G networks that can be investigated with OCUDU and MODRAD-SC.

\textit{Multi-user and massive MIMO}: 
    Multi-user and massive MIMO have become indispensable technologies in 4G and 5G systems \cite{3gpp38211}. The planned roadmaps at SRS and MB include the support of up to 64T64R massive MIMO systems for OCUDU and a massive MIMO MODRAD, respectively. 
    The Massive MIMO MODRAD will be a high-performance, 32-port TDD transceiver system designed for the $3.4$ to $3.8$ GHz frequency range, offering
    an average and peak output power of 23 dBm and 32 dBm, respectively, per port, $400$ MHz bandwidth and phase-coherent ports for advanced beamforming in 5G and 6G applications. The SDR mode will support real-time IQ sample streaming and deep memory playback via MATLAB or Python over 25G Ethernet.   
    This will enable an open environment for the investigation of, e.g., different O-RAN split options, BF, and channel estimation techniques with practical network infrastructure. Experiments can address standard-compliant methods in real-time and, using OCUDU and MODRAD-SC in SDR mode, techniques not included in the standards for pre-6G trials.

\textit{Cell-free MIMO}:
    Cell-free MIMO represents a very promising technology for reliable, low-latency, and high-bitrate connections thanks to its distributed antenna infrastructure \cite{ngo2024ultradense}. It is seen as a promising wireless network technology that is capable of handling the connectivity requirements in 6G. In a user-centric cell-free MIMO network, each terminal is connected to the part of the distributed antenna infrastructure that has the best radio channel characteristics, the so-called user-centric \textit{cluster}. Hence, a cell-free MIMO system would contain multiple spatially distributed MODRAD-SCs. Phase-accurate synchronization of the O-RUs is required to enable coherent transmission in the downlink, which can potentially be realized by wired and over-the-air (OTA) synchronization techniques (e.g., eCPRI, White Rabbit, PTP, reciprocity calibration). For the realization of an O-RAN cell-free MIMO system, the MODRAD-SCs would exchange information with one or multiple O-CU/O-DU instances responsible for cluster-level processing.
    
    %In an O-RAN cell-free MIMO system, the cluster formation will be realized by exchanging information such as channel characteristics and time-frequency resource availability between the MODRAD-SCs (in O-RAN mode), decentralized units and the Near-Real-Time RIC.

\textit{Waveform development}: In SDR mode, the MODRAD-SC can be used for the investigation of potential future 6G technologies that are not (yet) compliant with the O-RAN and 5G standards. One important research direction is \textit{advanced waveform design} considering, e.g., Orthogonal Time Frequency Space (OTFS), single carrier (SC) and enhanced Orthogonal Frequency-Division Multiplexing (OFDM) modulation. The waveforms can be evaluated using MODRAD-SCs as one or multiple transmitters and receivers. Such use case could be enabled in OCUDU through the creation of physical-layer hooks (APIs), e.g., by allowing to replace specific blocks of its current 3GPP-compliant signal processing infrastructure.

\textit{Sensing and localization}: Another use case combining OCUDU and MODRAD-SC in an SDR fashion is the experimental validation of integrated sensing and communication (ISAC) and RF imaging methods. Multiple MODRAD-SCs can serve as transmitter and receiver, while OCUDU can handle the resource management and the fusion of received data at different Rx MODRAD-SCs. While the observation receiver (ORx) ports at MODRAD-SC are generally used to capture and maintain a ``clean'' version the transmitted signal, in an ISAC use case they can also be used as reference to compute the sensed environment from the reflected received signal at the Rx ports.
Another research topic in this direction is user localization beyond FR1, which is currently under development by SRS for OCUDU, focused on an angle-based approach in the FR2 band.

\textit{NTN signaling}: MODRAD-SC addresses requirements such as timing constraints and REQ-RF-160 for openAMIP compatibility. Combining this with OCUDU's NTN features opens the door for interesting experiments in the TN/NTN coexistence and joint management and optimization (e.g., conditional handover). Considering the integration of TN-NTN coexistence in 6G, spectrum management becomes increasingly complex, lacking interoperability mechanisms. TN-NTN coordination can be investigated with OCUDU providing both the TN and NTN RANs, leveraging its inherent support for AI-driven monitoring and control, where MODRAD-SC can be used as O-RUs in both networks. 

\textit{FR1/FR3 coexistence}: The introduction of FR3 spectrum (7–15 GHz), as a 6G candidate band is essential for future capacity and coverage goals. Similar to TN-NTN coexistence, the FR1 and FR3 networks should be managed jointly to optimize their simultaneous operation. A potential system includes different MODRAD-SCs operating in FR1 and FR3, respectively, connected to one or several interacting O-CU/O-DU instances. The native RIC support offered by OCUDU can be exploited to dynamically manage the connection of the UEs to either FR1 or FR3, based on their location, band support, quality of service requirements, and load of each network. 

\textit{AI-driven network management}: AI will undoubtedly be a driving force in automated management and optimization of 6G networks for a myriad of different use cases (a few of which have been already mentioned throughout the text). When using OCUDU and with MODRAD for real-time experimentation (e.g., in a split 7.2 setup), the RICs can be used for dynamic RAN monitoring and control, for instance for energy-efficiency or performance optimization goals (e.g., load balancing, switching off or scaling RAN resources depending on current traffic needs, etc.). Moreover, for simulation-based research, combining OCUDU with MODRAD in SDR mode provides a solid experimental network to generate measurements that can be later used to train and test advanced AI-based methods and algorithms for network management and optimization. A relevant example is the generation of channel models and datasets to be latter exploited for AI-riven channel estimation, BF or radio resource allocation.

%{\BLUE [@Oriol: Can we write something meaningful here or should we remove this point? I tried it with a short and quite generic description.} In SDR mode, artificial intelligence (AI)-based methods and algorithms can be tested in an experimental network with measurement data. Compared to channel models and datasets, specific channel environments can be experimentally measured and used for AI-driven network management and optimization. The AI-based methods include, among others, channel estimation, BF, and radio resource allocation. 

%%%
%%%
%%%
%%%
\begin{figure}[t!]
	\centering
	\includegraphics[width=\linewidth]{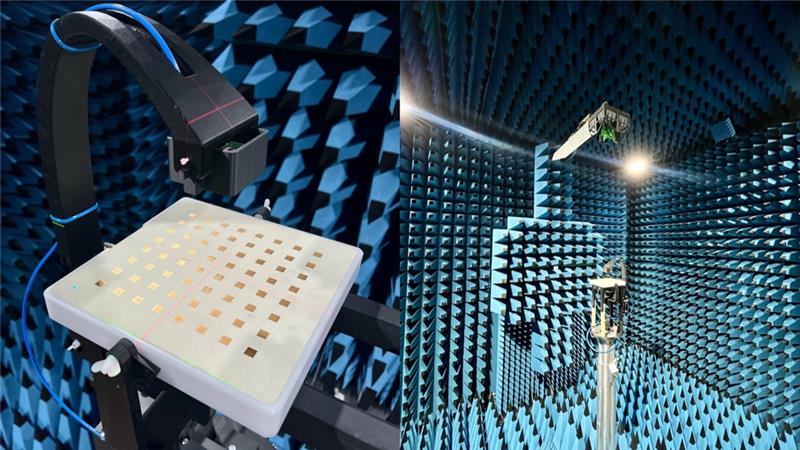}
	\vspace{-6pt}
	\caption{The FR3 NFED-RIS in a measurement chamber.}
	\label{fig_RIS_chamber}
    \vspace{-6pt}
\end{figure}
\subsection{FR3 Hybrid Beamforming with an NFED-RIS}
Higher carrier frequencies $f_c \in \left[ 7.125, 24.25 \right]$ GHz in frequency range 3 (FR3) are seen as potential frequency bands in 6G, in addition to sub-6 GHz, to enable high-bandwidth communication. To obtain the same aperture and coverage as a sub-6 GHz base station (BS), which allows the reuse of existing cell sites, a larger number of antennas (increased by factor 4 to 16) is required in FR3 because of the smaller wavelengths. Fully digital BF with a large number of active antennas results in high costs, large power demand and computational complexity. And even hybrid digital-analog approaches with an analog part in the RF and a digital part in the baseband domain \cite{molisch2017hybrid} are problematic at high carrier frequencies for the aforementioned reasons. 
One potential architectural solution to overcome these issues is the deployment of a near-field fed RIS, where a feeder array with a few active antennas illuminates the RIS in the near-field that is equipped with a larger number of passive antenna elements (phase shifters). This concept has been theoretically investigated in \cite{jamali2020intelligent, tiwari2025efficient}, while a practical realization by MB is presented in \cite{osterland2025evaluation, osterland2026evaluation}, which is also shown in Fig. \ref{fig_RIS_chamber} in a measurement chamber for far-field characterization.

Like in \cite{osterland2025evaluation, osterland2026evaluation}, we will consider the RIS being an $8 \times 8$ array equipped with 2-bit phase shifters resulting in phase shifts of $0^{\circ}$, $90^{\circ}$, $180^{\circ}$ and $270^{\circ}$, respectively. 
The performance of 2-bit phase shifters has been shown to be a good trade-off between performance (nearly optimal compared to continuous phase shifters) and complexity \cite{tiwari2025efficient}.
The phase shifters at the RIS steer the beam reflected by the RIS in the far field, and each phase shift configuration is associated to a beam ID (corresponding to a given beam direction in azimuth and elevation angle). 

\begin{figure}[t!]
	\centering
    \includegraphics[height=3.22cm]{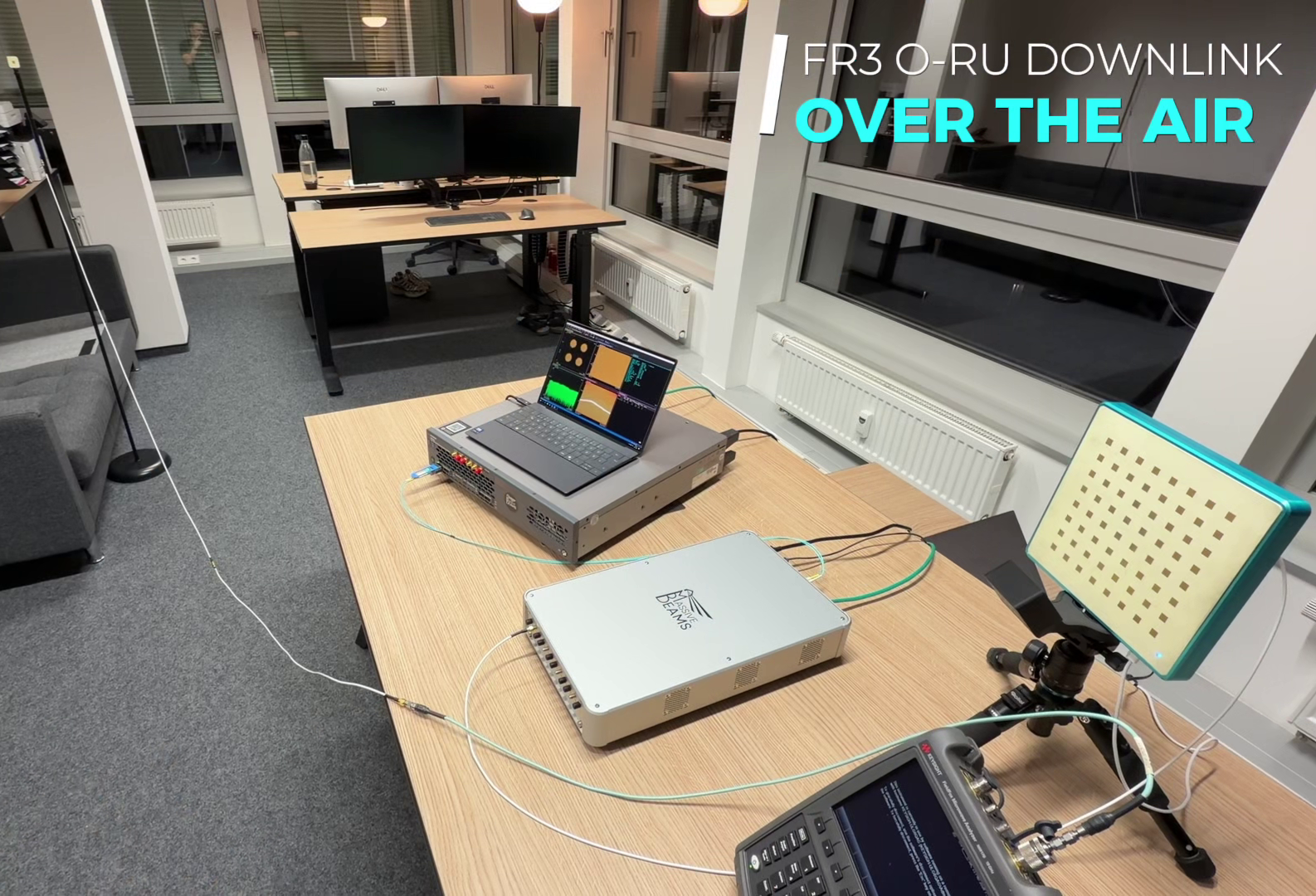}
	\includegraphics[height=3.22cm]{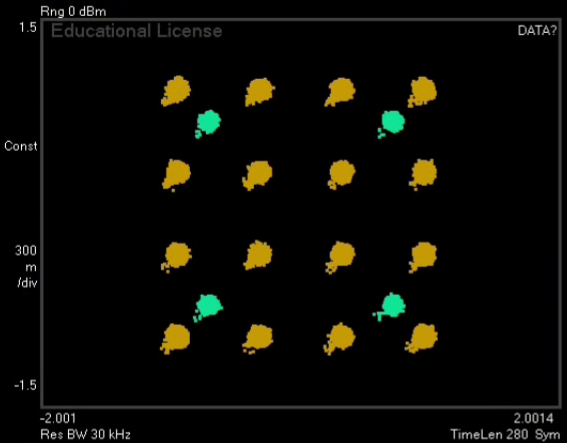}
	\vspace{-5pt}
	\caption{Downlink over-the-air transmission setup and symbol estimates at the receiver.}
	\label{fig_ota_measurements}
    \vspace{-5pt}
\end{figure}

In an O-RAN system, the beam IDs are transmitted from the O-DU to the O-RU over the fronthaul on the control plane (C-Plane).
In particular, the Section Type 1 for the C-Plane contains the field ``beamID'' (beam identifier) with a size of 15 bits, as defined in \cite{ETSI_TS103859_v1201}. 
Notice that there are $4^{64}\approx 3.4\times 10^{38}$ different phase shift configurations for a RIS with an $8 \times 8$ antenna array and 2-bit phase shifters. It is clear that not all configurations are meaningful since we cannot transmit beams across the complete angular domain (i.e., $360^{\circ}$ in both azimuth and elevation).
Instead, we assume that the NFED-RIS covers a region spanning 
$120^{\circ}$ in azimuth and $60^{\circ}$ in elevation. Furthermore, we construct the BF codebook such that the beams are separated by $3^{\circ}$ in both azimuth and elevation, resulting in a BF codebook of $800$ beam IDs. Hence, we need $10$ bits to assign a unique beam ID to each of the BF directions, fully in line with the length of the ``beamID'' field on the C-Plane.
The implementation of FR3 BF with the OCUDU, MODRAD-SC and an NFED-RIS is currently developed in the EU funded project 6G-LEADER (see \cite{6GLEADER_D71_2025}). The OCUDU transmits the ``beamID`` to the MODRAD-SC on the C-Plane, and the MODRAD-SC extracts the ``beamID`` from the received messages. The corresponding codebook, mapping ``beamID``s to phase shift configurations, is shared between OCUDU and MODRAD-SC. 
Fig. \ref{fig_ota_measurements} shows an over-the-air measurement setup at $f_c = 7.2$~GHz with a Keysight FieldFox as receiver connected to a single patch antenna. The green and orange symbol estimates correspond to pilot and 16-QAM data symbols, respectively. The beam shift configuration is optimized so that with the detected data symbols a small error vector magnitude is obtained.

Note that this  implementation can be applied and  extended to other beamformers. For example, OCUDU and MODRAD-SC in combination with holographic MIMO (HMIMO) surfaces would follow a similar approach. In a potential HMIMO setup, the MODRAD-SC is connected to an HMIMO array comprised of $M_m$ microstrips, each of which consists of $M_a$ metamaterial antennas. Then, each MODRAD-SC antenna port is connected to one microstrip to steer the RF signal, resulting in a total number of $M_m M_a$ antennas. OCUDU is responsible for providing the steering signals and configurations to the MODRAD-SC.  

% \begin{figure}[t!]
%     \captionsetup[subfigure]{font=footnotesize}
%     \centering
%     \begin{subfigure}[b]{0.485\linewidth}
%         \centering
%         \includegraphics[height=3.22cm]{figures/DL_setup_01.PNG}
%         \caption{}
%         \label{fig_setup}
%     \end{subfigure}
%     \hfill
%     \begin{subfigure}[b]{0.485\linewidth}
%         \centering
%         \includegraphics[height=3.22cm]{figures/constellation_01.png}
%         \caption{}
%         \label{fig_constellation}
%     \end{subfigure}
%     \caption{Downlink over-the-air transmission setup and symbol estimates at the receiver.}
%     \label{fig_ota_measurements}
% \end{figure}

\section{Conclusions} \label{sec_conclusions}

%{\RED [I'm not sure if we'll have space for "proper" results in case there are any (e.g., even a photo), but we need to highlight the potential of the combined OCUDU + MODRAD to be a key platform upon which execute experimental research for 6G. I've added a (very poor) first attempt.]}{\BLUE [@Oriol: Yes, due to the lack of results, I think we should focus on OCUDU + MODRAD as a key platform for experimental research for 6G. I made some small changes to your version, and I think it is more or less what we can conclude. But please feel free to make changes and add text.]}

The presented platform, consisting of SRS's OCUDU stack and MB's MODRAD-SC, is both versatile and very performant, yet its combination offers a very powerful experimental vehicle, enabling the practical evaluation of both standard-aligned and (custom) pre-6G technologies. OCUDU's openness and ambitions, including a massive development roadmap, combined with the modularity and richness of MODRAD-SC, results in an O-RAN-based solution offering an unparalleled capacity for practical research on advanced wireless communications. Its potential covers extremely relevant topics towards 6G, such as TN/NTN or FR1-FR2-FR3 coexistence and joint optimization, waveform development, multi-user and massive MIMO, and AI-driven radio/system management and optimization. The openness of the solution eases its adoption and the replicability of experiments, while addressing the requirements of academia, government, and industry for a reliable and future-proof O-RAN system for practical evaluation and real-world deployment.

\bibliographystyle{IEEEtran}
\bibliography{modrad_white_paper}

\end{document}